\documentclass[aps,pre,twocolumn,superscriptaddress,10pt]{revtex4-2}
\usepackage[utf8]{inputenc}
\usepackage{color}
\usepackage{amsmath}
\usepackage{amssymb}
\usepackage{xcolor} 
\usepackage{graphicx}
\usepackage{esint}
\usepackage{hyperref}
\usepackage{appendix}
\usepackage{comment}
\makeatletter
\@ifundefined{textcolor}{}
{%
\definecolor{red}{rgb}{0.75, 0.1, 0.1}
 \definecolor{BLACK}{gray}{0}
 \definecolor{WHITE}{gray}{1}
 \definecolor{GREEN}{rgb}{0,1,0}
 \definecolor{BLUE}{rgb}{0,0,1}
 \definecolor{CYAN}{cmyk}{1,0,0,0}
 \definecolor{MAGENTA}{cmyk}{0,1,0,0}
 \definecolor{YELLOW}{cmyk}{0,0,1,0}
}

\makeatother

\begin{document}
\title{Accelerating Quantum Relaxation via Temporary Reset: A Mpemba-Inspired Approach}
\author{Ruicheng Bao}
\email{Contact author: ruicheng@g.ecc.u-tokyo.ac.jp}
\affiliation{Department of Chemical Physics \& Hefei National Laboratory, University
of Science and Technology of China, Hefei 230088, China}
\affiliation{Department of Physics, Graduate School of Science,
The University of Tokyo, Hongo, Bunkyo-ku, Tokyo 113-0033, Japan}
\author{Zhonghuai Hou}
\email{Contact author: hzhlj@ustc.edu.cn}
\affiliation{Department of Chemical Physics \& Hefei National Laboratory, University
of Science and Technology of China, Hefei 230088, China}
\begin{abstract}
Slow relaxation processes spanning widely separated timescales pose
fundamental challenges for probing steady-state properties and engineering
functional quantum systems, such as quantum heat engines and quantum
computing devices. We introduce a protocol that enables significant acceleration of relaxation in general Markovian open quantum systems by temporarily coupling the system to a reset channel, inspired by the Mpemba effect. Crucially, this acceleration persists even when the slowest decaying Lindbladian modes form complex-conjugate
pairs. Unlike previous approaches, which typically target a single mode, our protocol may suppress multiple relaxation modes simultaneously. This framework provides a versatile and experimentally feasible tool
for controlling relaxation timescales, with broad implications
for quantum thermodynamics, computation, and state preparation.

\end{abstract}
\maketitle
\textit{Introduction.}---Relaxation processes in open quantum systems,
wherein a driven or interacting system settles into an equilibrium
or a nonequilibrium steady state, are of fundamental importance in
nonequilibrium physics and for practical quantum technologies. The
timescales of such relaxations often determine the feasibility and
performance of quantum devices. For example, in quantum thermodynamics,
the power output of cyclic heat engines or transport junctions can be limited by how rapidly the working substance relaxes to its steady state in each cycle \cite{17PRL_QHE,22PRR_QHE}. In addition, fast relaxation
facilitates ground state laser cooling \cite{00lasercooling,20lasercooling,21lasercooling}
and the reliable preparation of quantum states \cite{PRApreparation}.
Faster relaxation can also lead to more efficient quantum algorithms
\cite{Schulman2005}, particularly in dissipative computational architectures
where the output is encoded in the system's stationary state \cite{Schulman2005,Verstraete2009}.
Slow relaxation inevitably allows other unwanted dissipative dynamics
to consume substantial time resources, compromising the efficiency
of the final stabilized state \cite{21PRL_Qmpemba,25PRL_Accelerate}.
In contrast, slow relaxation is sometimes desirable---such as when
the computational task targets metastable states \cite{daley2011metastable,allcock2021metastable,Yang2022meta,25PRAmeta},
whose prolonged lifetimes are essential for robust information storage
or approximate optimization.

Researchers have developed various techniques to accelerate relaxation
toward both thermal equilibrium \cite{19PRL_shortcut} and nonequilibrium
steady states \cite{23PRR_EPaccelerate,24PRXQuantum_tradeoff,25PRL_Accelerate}.
Most of these methods, however, are restricted to specific systems
and rely on intricate external controls. For instance, some approaches
require auxiliary Hamiltonians or full knowledge of the system Hamiltonian,
limiting their applicability to simple setups. A more general and
elegant strategy is to accelerate relaxation by tailoring the initial
state distribution, without the need for continuous external driving.
A particularly intriguing example of this approach is inspired by
the so-called Mpemba effect \cite{69Mpemba}, a counterintuitive relaxation
anomaly in which a hotter system can cool faster than a colder one.
This phenomenon implies the existence of an optimal initial state
that minimizes the relaxation timescale. Theoretical frameworks for
identifying such optimal states have been proposed in both classical
\cite{lu2017nonequilibrium,klich2019mpemba,kumar2020exponentially,santos2020mpemba,gal2020precooling,Busiello_2021,Bao23prr,PREBrownian2023,RazboundaryPRL,24ObsevableMpemba,Shapira2024,tanthermomajorization,ohga2024microscopic}
and quantum systems \cite{21PRL_Qmpemba,22PRA_Mpemba,ares2023entanglement,23PRLQuantumdotMpemba,Joshi2024,Liu2024,Nava2024,multimpemba24,Rylands2024,Wang2024,Gooldprl24},
see also \cite{teza2025speedups,ares2025quantum} for recent reviews.
However, existing quantum implementations typically impose strict
constraints, such as assuming pure initial states or requiring the
second-largest eigenvalue of the Lindbladian to be real. Moreover, careful initial-state design is delicate, since it relies on fine-tuning of control parameters and detailed knowledge of the initial state and system dynamics, which are typically not known \textit{a priori} in complex quantum systems. These issues limit
their applicability in general settings. Thus, a general and practical method for inducing faster relaxation in open quantum systems, which is robust with respect to different initial states, remains elusive. 

In this study, we contribute to addressing these challenges by proposing an experimentally
feasible protocol that can both accelerate and decelerate general
quantum relaxation processes  {from any initial state, by resetting the system to a specified (possibly mixed) target state \cite{18PRE_spectral,18PRB_Qreset,22SciRep_Qreset}. Here, this type of quantum reset operation can be realized through engineered dissipation, i.e., quantum reservoir engineering \cite{96PRL_zoller,zoller2008quantum,Verstraete2009,harrington22review}, and should be distinguished from the conventional reset typically used for qubit initialization in quantum computation \cite{22PRL_3thlaw,Basilewitsch_2017,Basilewitsch_2019,19PRA_qbreset,21PRR_qbreset,23PRL_Ruoyu,25pnas_ruoyu,Aamir2025}.} In our protocol, we introduce a finite-duration reset phase into the open-system dynamics by temporarily coupling the system to a reset
channel. We show that this protocol can significantly accelerate
relaxation, even when starting from mixed initial states and in the
presence of complex decay modes. Remarkably, our method allows for
the simultaneous suppression of multiple relaxation modes---a capability
absent in previous approaches. Moreover, selective application of the protocol enables the system of interest to remain in metastable states for extended periods, offering
a systematic route to suppress relaxation.

\textit{Setup.}---We consider a general Markovian open quantum system defined
on a Hilbert space $\mathcal{H}$ of dimension $d$, whose dynamics are governed by the Lindblad-Gorini-Kossakowski-Sudarshan
(LGKS) master equation $d\rho/dt=\mathcal{L}(\rho),$ where the Lindbladian
superoperator $\mathcal{L}$ is given by
\begin{equation}
\mathcal{L}(\rho)\equiv-i[H,\rho]+\sum_{i}\left[J_{i}\rho J_{i}^{\dagger}-\frac{1}{2}\{J_{i}^{\dagger}J_{i},\rho\}\right].\label{Lindbladian}
\end{equation}
Here, $H$ is the Hamiltonian of the system and the jump operators
$J_{i}$ describe the dissipative coupling to the environment. 
The evolution can be analyzed via the spectral decomposition
of $\mathcal{L}$, whose eigenvalues are denoted $\lambda_{k}$ and
ordered such that $0 \geq \text{Re}(\lambda_k)\geq \text{Re}(\lambda_{k+1})$. 
Assuming that $\mathcal{L}$ is diagonalizable,
let $R_{k}$ and $L_{k}$ denote the corresponding right and left
eigenmatrices, satisfying $\mathcal{L}(R_{k})=\lambda_{k}R_{k}$ and
$\mathcal{L^{\dagger}}(L_{k})=\lambda_{k}^{\star}L_{k},\ k=1,...,d^{2}$,
with biorthogonal normalization \cite{tutorial_24prxquantum}
\begin{equation}
\text{Tr}(L_{k}^{\dagger}R_{h})=\delta_{kh}.\label{normalized}
\end{equation}
The dual superoperator $\mathcal{L}^{\dagger}$ governs the Heisenberg-picture
dynamics of observables:
{\begin{equation*}
    \mathcal{L}^{\dagger}(O)=i[H,O]+\sum_{i}\left[J_{i}^{\dagger}OJ_{i}-\frac{1}{2}\{J_{i}^{\dagger}J_{i},O\}\right].
\end{equation*}}
Given an initial state $\rho_{0}$, the system state at time $t$
evolves as
\begin{equation}\label{spectral_original}
\rho(t)=e^{t\mathcal{L}}[\rho_{0}]=\rho_{\rm ss}+\sum_{k=2}^{d^{2}}c_ke^{\lambda_{k}t}R_{k},
\end{equation}
where $c_{k}\equiv\text{Tr}(L_{k}^{\dagger}\rho_{0})$. The unique
stationary state of the open quantum system $\rho_{\rm ss}$ is given
by $\rho_{\rm ss}=\lim_{t\rightarrow\infty}\rho(t)=R_{1}$, assuming $\lambda_{1}=0$
is non-degenerate. In the long-time limit, the relaxation is dominated
by the slowest decaying mode, and the deviation from stationarity
obeys $||\rho(t)-\rho_{\rm ss}||\sim\exp\left(-|\text{Re}\lambda_{2}|t\right)$.

\textit{Reset protocol to accelerate or decelerate relaxation processes}.---Recognizing
that the slowest decaying mode governs the relaxation timescale, we
propose a quantum reset protocol that can suppress or promote its excitation from
a general initial state via a finite-duration reset phase, thereby enabling exponential acceleration of relaxation. Specifically, we
let the system evolve under modified dynamics for a finite time $t_{s}$,
during which it is stochastically reset to a chosen state $\rho_{\delta}$
at random times, with events following a Poisson process of rate $r$. After time $t_{s}$, the reset channel is turned off, and the system
resumes evolving under the original Lindbladian $\mathcal{L}$. {That is, the system density matrix $\rho(t)$ evolves under the reset channel for $t\in [0,t_s]$ and decouples from the channel for $t>t_s$}. The
modified dynamics during the reset phase $t\in[0,t_s]$ is governed by a new Lindbladian
$\mathcal{L}_{r}$ \cite{18PRE_spectral}, defined as 
\begin{equation}\label{modified_L}
\mathcal{L}_{r}(\rho):=\mathcal{L}(\rho)+\mathcal D_r(\rho)=\mathcal{L}(\rho)+r\text{Tr}(\rho)\rho_{\delta}-r\rho,
\end{equation}
where $\mathcal D_r(\rho)= r[\text{Tr}(\rho)\,\rho_\delta-\rho]$. Expressing $\rho_{\delta}=\sum_{\alpha}p_{\alpha}|\psi_{\alpha}\rangle\langle\psi_{\alpha}|$, $\mathcal{L}_r$ is equivalent to introducing jump operators $J_{i,\alpha}^{r}=\sqrt{rp_{\alpha}}|\psi_{\alpha}\rangle\langle\phi_{i}|$
into $\mathcal{L}$, where $|\phi_{i}\rangle$
form an auxiliary complete orthonormal basis (see End Matter). The term $r\text{Tr}(\rho)\rho_{\delta}$ contributes only to the stationary mode
and vanishes on all traceless eigenmodes $R_{i}$ with $i\geq2$,
which satisfy $\mathrm{Tr}(R_{i})=0$ due to $\mathrm{Tr}(L_{1}^{\dagger}R_{i})=0$
and $L_{1}^{\dagger}=\mathbb{I}$ {[}cf. (\ref{normalized}){]}. 
Consequently, the modified Lindbladian acts as 

\begin{equation}\label{spectral_shift}
\mathcal{L}_{r}(R_{i})=(\lambda_{i}-r)R_{i},\ i\in\{2,...,d^{2}\}.
\end{equation}
{This induces a uniform spectral shift by $-r$ for all non-stationary eigenmodes, leaving the eigenmatrices ($R_i$, $i \geq 2$) themselves unchanged. In contrast, the stationary mode $R_1$ retains eigenvalue zero but is modified under reset.} Let $\rho^{r}(t):=e^{t\mathcal{L}_{r}}\rho_{0}$ be
the state at time $t$ under reset dynamics. Its spectral decomposition
reads:
\begin{align}\label{spectral_modified}
\rho^{r}(t)&=  e^{t\mathcal{L}_{r}}\rho_{\rm ss}+\sum_{k=2}^{d^{2}}c_k e^{(\lambda_{k}-r)t}R_{k}\nonumber \\
&:=  \rho_{\rm ss}+\sum_{k}c_{k}^{r}(t)e^{\lambda_{k}t}R_{k},\ (0\leq t\leq t_s)
\end{align}
where {$c_{k}^{r}(t)$}
are time-dependent mode amplitudes. Explicitly, these modified coefficients
are given by (End Matter)
\begin{equation}\label{modified_coeff}
c_{k}^{r}(t)\equiv\left[c_{k}-\frac{r\cdot d_{k}}{r-\lambda_{k}}\right]e^{-rt}+\frac{r\cdot d_{k}}{r-\lambda_{k}}e^{-\lambda_{k}t},
\end{equation}
where $d_{k}\equiv\text{Tr}(L_k^{\dagger}\rho_{\delta})$ is the overlap coefficient of $\rho_{\delta}$. Notably, Eq. \eqref{spectral_modified} can be generalized to the dynamics of observables \cite{supplemental_material}, which may be more convenient to measure. 

After the reset phase ($t>t_{s}$), the system returns to the original Lindbladian evolution. {The dynamics for $t>t_s$ then reads:
\begin{equation}
\rho^{r}(t) =  \rho_{\rm ss}
  +\sum_{k}c_{k}^{r}(t_{s})e^{\lambda_{k}t}R_{k}. \quad (t>t_s)\label{wholedynamics}
\end{equation}

Importantly, the modified spectral decomposition, Eq.~\eqref{wholedynamics}, takes the same form as the original one, Eq.~\eqref{spectral_original}, which allows us to directly compare the relaxation dynamics with and without reset. From this comparison, it is clear that for $t>t_s$ our protocol is equivalent to the original Lindbladian evolution, but with different overlap coefficients,
\begin{equation}
    \rho(t)=\rho_{\rm ss}+\sum_{k=2}^{d^{2}}c'_k e^{\lambda_{k}t}R_{k}. \quad (t>t_s)
\end{equation}
Here, $c_k'\equiv c^r_k(t_s)$ are determined by the reset protocol. 

Thus, our protocol controls relaxation dynamics by tuning the overlap coefficients $c_k$, without requiring any special preparation of the initial state. By choosing an appropriate $\rho_{\delta}$ and varying $r$ and $t_s$, one can tune the amplitudes of multiple relaxation modes. In particular, ensuring $|c_2^r(t_s)|<|c_2|$ ($>|c_2|$) suppresses (enhances) the dominant mode, which is sufficient for acceleration (deceleration), ultimately bringing the system closer to (farther from) stationarity~\cite{lu2017nonequilibrium,22PRA_Mpemba}. Specifically, the dominant mode is fully eliminated and exponential speed-up occurs when $|c_2^r(t_s)|=0$; this, however, requires fine-tuning and prior knowledge and is therefore challenging to realize in large systems. We thus focus on the weaker but practically relevant scenario $|c_2^r(t_s)|<|c_2|$.

\textit{Sufficient conditions for suppression or promotion of relaxation modes.---}
We derive a sufficient condition \cite{supplemental_material},
\begin{equation}\label{Mpemba_condition}
    {\rm Re}(c_2^*d_2)< |c_2|^2,
\end{equation}
under which there always exists a threshold $t_c>0$ such that, for any $r>0$ and any $t_s\in(0,t_c]$, one has $|c_2^{r}(t_s)|<|c_2|$. That is, the relaxation is accelerated for all $t_s$ in this range. If the condition is violated, there exists another threshold $t_c'>0$ with $|c_2^{r}(t_s)|\geq|c_2|$ for $t_s\in(0,t_c']$, leading to deceleration. However, Eq. \eqref{Mpemba_condition} is not necessary for acceleration, since one may still have $|c_2^{r}(t_s)|<|c_2|$ for $t_s>t_c'$, when it is not satisfied.

Several remarks are in order. First, the condition applies both when $\lambda_2$ is real and when it forms a complex-conjugate pair. In the latter case, the long-time dynamics takes the form
\begin{equation}
\rho(t)=\rho_{\rm ss}+e^{\lambda_{2}t}c_{2}R_{2}+e^{\lambda_{2}^{*}t}c_{2}^{*}R_{2}^{\dagger}+\mathcal{O}(e^{\lambda_3t}),
\end{equation}
where $c_{2}=\text{Tr}(L_{2}^{\dagger}\rho_{0})$ and $c_{2}^{*}=\text{Tr}(L_{2}\rho_{0})$
is its complex conjugate. Most existing strategies cannot simultaneously
suppress both components of such complex relaxation modes. Our protocol, however, naturally addresses this limitation.
The modified coefficients under reset are
\begin{subequations}
     \begin{align}
c_{2}^{r}(t)&=\left[c_{2}-\frac{r\, d_{2}}{r-\lambda_{2}}\right]e^{-rt}+\frac{r\, d_{2}}{r-\lambda_{2}}e^{-\lambda_{2}t},\\
c_{2}^{r,*}(t)&=\left[c_{2}^{*}-\frac{r\, d_{2}^{*}}{r-\lambda_{2}^{*}}\right]e^{-rt}+\frac{r\, d_{2}^{*}}{r-\lambda_{2}^{*}}e^{-\lambda_{2}^{*}t}.
\end{align}
\end{subequations}
These coefficients remain complex conjugates for all $t$, i.e., $[c_{2}^{r}(t)]^{*}=c_{2}^{r,*}(t)$. Eq.~\eqref{Mpemba_condition} then ensures that $|c_{2}^{r,*}(t_s)|=|c_{2}^{r}(t_s)|\leq |c_2|=|c_2^*|$ for a moderate $t_s$.

Second, Eq.~\eqref{Mpemba_condition} is a weak condition satisfied by many choices of $\rho_{\delta}$. In particular, it does not require that $\rho_{\delta}$ be closer to the steady state than $\rho_0$, nor that the overlap coefficient $|d_2|$ be smaller than $|c_2|$. For example, when $\lambda_2$ is real, if ${\rm Re}(d_2)$ and $c_2$ have opposite signs, the condition is satisfied even when $|d_2|$ is arbitrarily large. In Example~2 we show that the condition is typically satisfied for a large fraction of randomly chosen initial states.

Third, neither $r$ nor $t_s$ needs to be precisely tuned in advance to achieve acceleration or deceleration. Provided that Eq. \eqref{Mpemba_condition} holds, any positive $r$ and moderately small $t_s$ are sufficient for acceleration, and $t_s$ can be further optimized \textit{a posteriori} based on the observed dynamics. Our protocol therefore provides a general and robust framework for controlling relaxation in open quantum systems, without the need for fine-tuning. 

Finally, it is straightforward to see that
\begin{equation}\label{Mpemba_condition_k}
    {\rm Re}(c_k^*d_k)< |c_k|^2
\end{equation}
is a sufficient condition for suppression of the $k$-th mode. Since multiple such conditions for different $k$ can be satisfied simultaneously, our protocol allows simultaneous suppression (or promotion) of multiple relaxation modes. In \cite{supplemental_material} we provide numerical evidence that multiple conditions can indeed be satisfied in parallel, using the model and protocol of Example 2.


So far we have focused on practical scenarios where fine-tuning is not possible and neither the system dynamics nor the initial state are known \textit{a priori}. If these constraints are lifted, as in most previous studies, our protocol can fully eliminate the dominant relaxation mode ($|c_2^r(t_s)|=0$) by choosing an optimal $t_s=t_s^*$. When $\lambda_2$ is real, the exact expression is \cite{supplemental_material,Busiello_2021}
\begin{equation}
t^*_{s}=\frac{1}{r-\lambda_{2}}\ln\!\left(1-\frac{c_{2}(r-\lambda_{2})}{r d_{2}}\right).\label{ts}
\end{equation}
A sufficient condition ensuring that $t^*_{s}\geq 0$ is $c_2/d_2\leq 0$. Further details, including the case of complex-conjugate pairs, are given in \cite{supplemental_material}.}

\textit{Example 1.}---To illustrate our protocol, we first consider a minimal
example: a two-level quantum system, which may represent a qubit or
a spin-$1/2$ particle. {While minimal, this example is crucial in many practical settings, as qubit reset often constitutes the bottleneck in large-scale quantum processes \cite{24PRXQuantum_qreadout,Aamir2025}.} We set the ground state energy $E_{0}=0$
and excited state energy $E_{1}=E$, such that the system Hamiltonian
reads $H=E|1\rangle\langle1|+\Omega(\sigma^{+}+\sigma^{-})$, where $\Omega$ is the intrinsic coherent coupling between the two
levels, $\sigma^{+}=|1\rangle\langle0|$ and $\sigma^{-}=|0\rangle\langle1|$. The system is coupled to a thermal environment through the
Lindblad jump operators $J_{0}=
\sqrt{\gamma_0}\sigma^+$ and
$J_{1}=\sqrt{\gamma_1}\sigma_{-}$. Here, $\gamma_{0}$ ($\gamma_1$) is the transition rate from state $|0\rangle$
to state $|1\rangle$ (or vice versa). Assuming the rates satisfy detailed
balance with inverse temperature $\gamma_{0}e^{-\beta_{\text{env}}E_{0}}=\gamma_{1}e^{-\beta_{\text{env}}E_{1}}$,
i.e., $\gamma_{0}=\gamma_{1}e^{-\beta_{\text{env}}E},$ the system
evolves to a unique equilibrium state $\rho_{\rm eq}=\frac{e^{-\beta_{\text{env}}H}}{\text{Tr}(e^{-\beta_{\text{env}}H})}$
irrespective of the initial condition. We initialize the system in
a general mixed state:
\begin{align*}
\hat{\rho}_{0} & =\left(\begin{array}{cc}
\frac{1}{1+e^{-\beta_{0}E}} & ke^{i\phi}\\
ke^{-i\phi} & \frac{e^{-\beta_{0}E}}{1+e^{-\beta_{0}E}}
\end{array}\right),\\
0\leq k & \leq\frac{\sqrt{e^{\beta_{0}E}}}{e^{\beta_{0}E}+1},\quad0\leq\phi\leq2\pi,
\end{align*}
where the off-diagonal terms $ke^{\pm i\phi}$ quantify initial quantum
coherence. {When $\Omega=0$,
all eigenmodes and eigenvalues of
the Lindbladian can be computed analytically \cite{supplemental_material}.} When
we apply our reset protocol with the reset state $\rho_{\delta}=|0\rangle\langle0|$,
corresponding to the ground state, the
two slowest decaying modes form a complex-conjugate pair:
\begin{equation}
c_{2}^{r}(t)=e^{-rt+i\phi}k,\quad c_{2}^{r,*}(t)=e^{-rt-i\phi}k.
\end{equation}
{These overlaps are exponentially suppressed compared with the original overlap, i.e. $c_{2}^{r}(t)=e^{-rt}c_2$ and $c_{2}^{r,*}(t)=e^{-rt}c_2^*$.}
{Clearly, the relaxation is accelerated for
any $r$ and $t_s$.} 

Since any $t_s$ can ensure acceleration, we choose $t_{s}$ such
that a faster decaying mode is eliminated.
Specifically,
solving $c_{4}^{r}(t_{s})=0$ gives such a $t_s$, whose expression is provided in \cite{supplemental_material}.
For a sufficiently large $r$, the slowest mode's amplitude becomes exponentially small after the reset phase. Alternative selections for $t_{s}$ can also achieve
accelerated relaxation, indicating that detailed dynamics knowledge
is not essential. Notably, if the initial inverse temperature satisfies
$\beta_{0}<\beta_{\text{env}}$, a scenario analogous to ``cooling'',
the ground state $|0\rangle\langle0|$ serves as a useful
$\rho_{\delta}$ regardless of the rate $r$. Conversely, when $\beta_{0}>\beta_{\text{env}}$,
resetting to the ground state decelerates relaxation, necessitating the choice $\rho_{\delta}=|1\rangle\langle1|$ for acceleration.
This example demonstrates that, by appropriately choosing
$\rho_{\delta}$, one can switch between reset-induced acceleration and deceleration of relaxation.

\begin{figure}
\begin{centering}
\includegraphics[width=1\columnwidth]{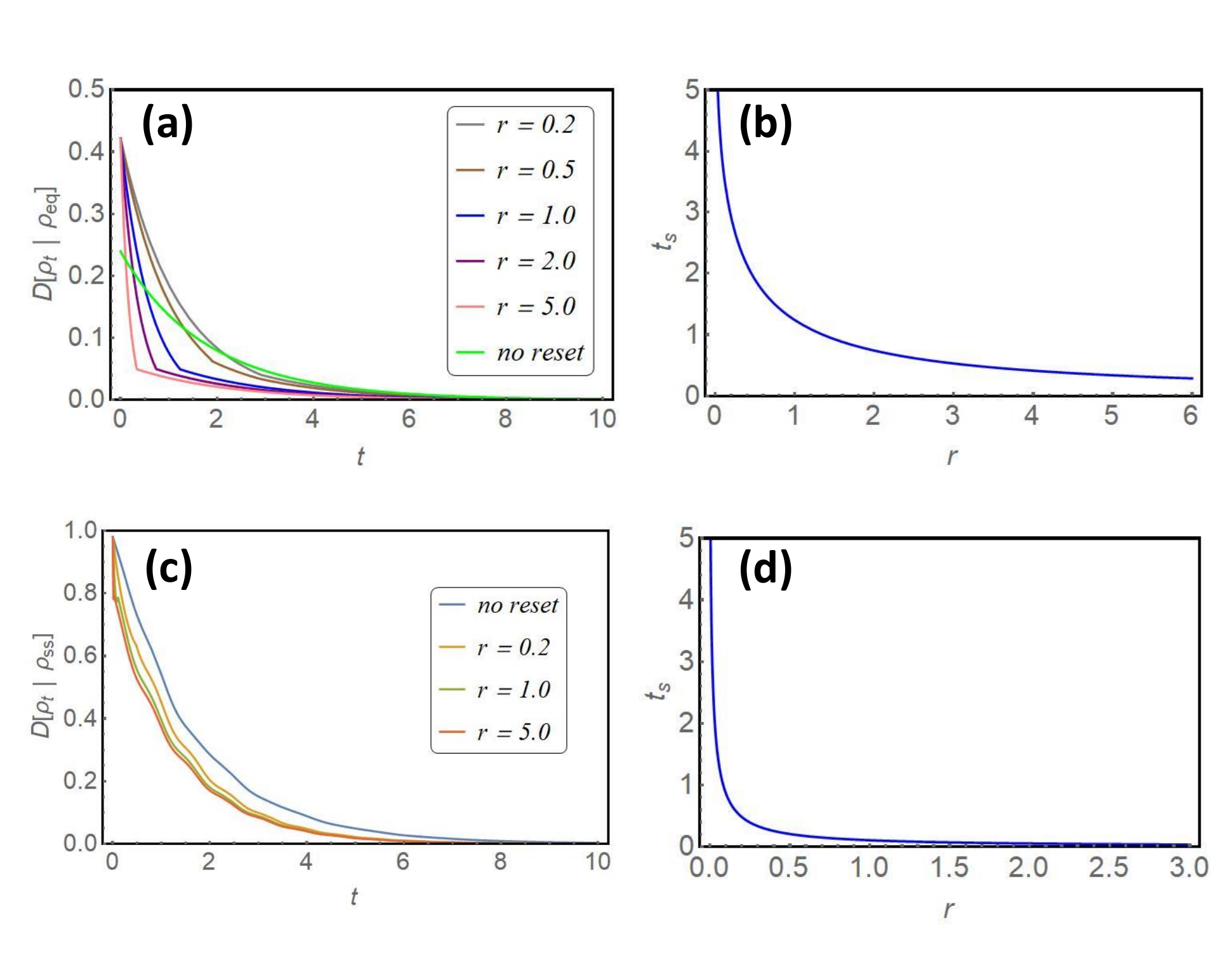}
\par\end{centering}
\caption{Relaxation dynamics of the two-state system. (a) and (c): Distances
between $\rho(t)$ and $\rho_{\text{eq}}$ as a function of time
$t$ with different resetting rates $r$. (b) and (d): The $t_{s}$ as a function of the resetting rate $r$. The parameters
are chosen as $\beta_{0}=2.0$, $k=0.32$ and $\phi=1$ with reset
protocol and $\beta_{0}=3.0,$ $k=0.21$ and $\phi=1$ without reset. The environment is at lower temperature $\beta_{\text{env}}=4.0$. $\gamma_1=1.0$.
For (a) and (b) $\Omega=0$, for (c) and (d) $\Omega=2.0$.}

\label{fig1}
\end{figure}

We next present numerical
results for illustration. We characterize the distance between the transient state
$\rho(t)$ and the stationary state $\rho_{\rm ss}$ using the standard trace distance $D[\rho(t)|\rho_{\rm ss}]:=\text{Tr}|\rho(t)-\rho_{\rm ss}|/2$
where $|A|:=\sqrt{A^{\dagger}A}$. Other measures, such as the $L_{\infty}$
norm $D_{\infty}[\rho(t)|\rho_{\rm ss}]:=\max_i|\lambda_i|$ [$\lambda_i$ is the $i$-th eigenvalue of $\rho(t)-\rho_{\rm ss}$], yield qualitatively similar outcomes. We first plot $D[\rho(t)|\rho_{\rm eq}]$ over $t$
for various rates $r$ in Fig. \ref{fig1} (a). {The parameters for $r>0$ cases are chosen so that they are initially farther from equilibrium compared to the $r=0$ case (see the caption of Fig. \ref{fig1}). The reset protocol substantially accelerates relaxation so that the former cases reach stationarity faster even if they are farther from stationarity initially.} We next plot $t_{s}$ as a function of $r$ in Fig. \ref{fig1} (b) and (d), which show that $t_{s}$ decreases monotonically as $r$
increases.  
Fig. \ref{fig1} (c) confirms accelerated
relaxation for $\Omega=2$, here measured by the $L_{\infty}$ norm to reduce fluctuations in the curves. {Notably, the steady state is not equilibrium when $\Omega>0$.} 

{\textit{Example 2.}---To illustrate the potential feasibility of our protocol in complex quantum systems, we consider another example, the dissipative transverse-field Ising model (TFIM), a paradigmatic system relevant to various quantum platforms. We numerically study an open dissipative TFIM of length $N$ with Hamiltonian
\begin{equation}  
H =-J\sum_{i=1}^{N-1}\sigma_i^z\sigma_{i+1}^z
     -g\sum_{i=1}^{N}\sigma_i^x , 
\end{equation}
and jump operators $J_{i,\downarrow}=\sqrt{\gamma}\,\sigma_i^-,
\quad
J_{i,\uparrow}  =\sqrt{\gamma e^{-\beta}}\;\sigma_i^+$. Here, $\sigma^{x,z}_i$ are Pauli operators, $\sigma_i^-=|0\rangle \langle 1|$ and $\sigma_i^+=|1\rangle \langle 0|$.
We fix $J=1$, $g=1.2$, $\gamma=0.5$, $\beta=1/k_BT=1$ throughout.
We choose $\rho_{\delta}=\mathbb{I}/d$ (with $d=2^N$).  
The initial state $\rho_0$ is chosen by normalizing $\rho_{\rm ss}+\alpha V_2$, with $\rho_{\rm ss}$ being the steady-state and $V_2$ being the second normalized eigenvector of the Lindbladian. A larger $\alpha$ implies that $\rho_0$ is farther from $\rho_{\rm ss}$. Here, $\rho_{\rm ss}$ is nontrivial, with nonzero entanglement. {Moreover, as shown in \cite{supplemental_material}, for a large fraction of randomly chosen pure initial states our protocol with $\rho_{\delta}=\mathbb{I}/d$ still yields substantial acceleration, further demonstrating its robustness.}
Parameters: $t_s=0.5\,\tau_2$ and $t_s=0.08\,\tau_2$ with $\tau_2=1/|\operatorname{Re}\lambda_2|$, the relaxation timescale, total duration $T=6\,\tau_2$, $\alpha=0.55/0.05$. Notably, $t_s$ is chosen arbitrarily, rather than being determined \textit{a priori}. 
The accelerated relaxation processes under different reset rate are shown in Fig. \ref{fig2}. We also examine cases with different parameters in \cite{supplemental_material}, where acceleration is consistently observed. In Fig. \ref{fig2} (b) and (d), genuine Mpemba effects are observed, where the system is first driven away from the steady state by the reset protocol compared to the no-reset case, and then reaches stationarity faster. The existence of such an effect broadens the range of usable reset states.

{We further verify in \cite{supplemental_material} that the same protocol significantly accelerates relaxation in another many-body setting, the Dicke model, again using $\rho_{\delta}=\mathbb{I}/d$.}

{Some remarks are in order regarding experimental implications. In practice, the precise optimal duration $t_s$ is generally unknown in advance, but as emphasized before, fine-tuning is unnecessary. The TFIM example directly supports this conclusion: any choice of $t_s$ shorter than the intrinsic relaxation time already leads to a clear acceleration.} 
Additionally, the reset channel in the example corresponds to a depolarizing channel, which is conceptually straightforward to implement, e.g., by applying an isotropic white-noise field or coupling an infinite-temperature reservoir. 

\begin{figure}
\begin{centering}
\includegraphics[width=1.0\columnwidth]{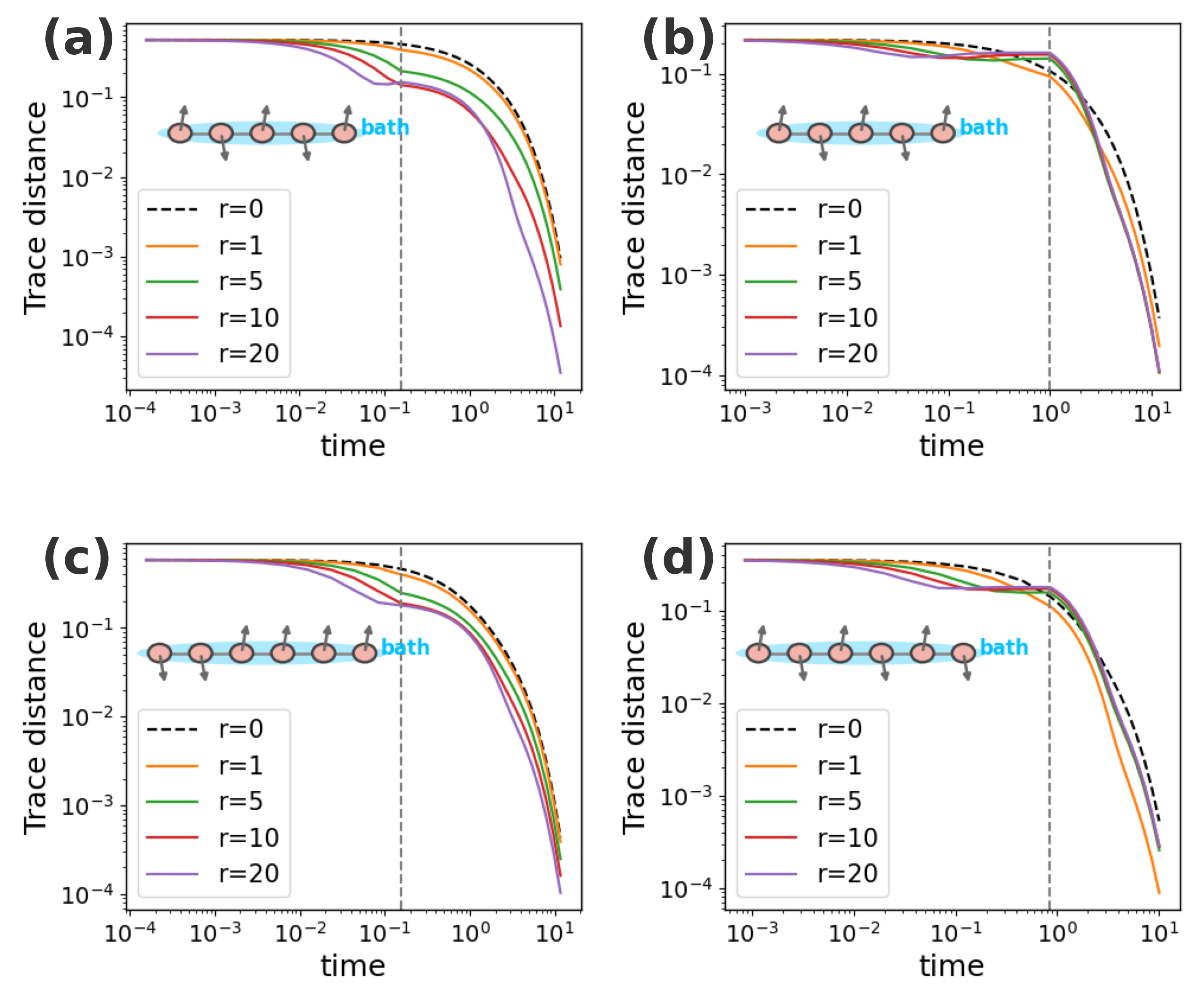}
\par\end{centering}
\caption{Acceleration of the relaxation of TFIMs. Reset rates $r$ are chosen as $0, 1.0, 5.0, 10.0, 20.0$. (a) $N=5$ ($d=32$), $\alpha=0.55$, $t_s=0.08\tau_2$. (b) $N=5$, $\alpha=0.05$, $t_s=0.50\tau_2$. (c) $N=6$ ($d=64$), $\alpha=0.55$, $t_s=0.08\tau_2$. (d) $N=6$, $\alpha=0.05$, $t_s=0.50\tau_2$.  The vertical gray lines mark $t_s$ at which the reset channel is removed. Inset: Schematic representation of the dissipative TFIM, with each spin coupled to a thermal bath.}
\label{fig2}
\end{figure}
}

{\textit{Experimental feasibility in general cases.}---The TFIM example illustrates the power of our protocol: the acceleration of relaxation via a depolarizing channel has the potential to be applied to complex quantum platforms. However, in general, it may be necessary to choose other $\rho_{\delta}$ to achieve acceleration. For few-body systems (such as single Rydberg atoms or other multi-level emitters), the protocol may be implemented with quantum reservoir engineering. Such techniques have become increasingly routine for tailoring dissipative dynamics in diverse quantum systems, including
atoms \cite{12PRL_bathengineering,21prl_atomreservoir,APL_reservoirengineering}, trapped ions \cite{96PRL_zoller,kienzler2015science}, superconducting circuits \cite{12PRL_bathengineering,harrington22review,22prl_superconducting}, and optomechanical
setups \cite{13PRL_optomech,15science_opto,17nphy_optomech}. 
For quantum many-body systems, an exact implementation via reservoir engineering generally requires high-order interactions. Such engineered $N$-body dissipators have been put forward theoretically \cite{weimer10rydberg_nature,20PRR_reservoir} and realized experimentally up to four-body interactions \cite{barreiro11nature}. 

A complementary route is to approximate the reset channel:  combine any available state-preparation protocol for $\rho_{\delta}$, whether reservoir engineering, measurement-based feedback, or other methods, with Trotterization \cite{16PRL_Trotter,21PRX_Trotter,22PRXQuantum_Trottersimu} (see End Matter). This strategy sidesteps the need to engineer reset jump operators directly and therefore avoids high-order couplings even in many-body systems. The main experimental challenge then reduces to preparing $\rho_{\delta}$.  Encouragingly, fast and high-fidelity state preparation has been reported across diverse platforms \cite{19PRL_BBprepare,22PRXQuantum_fastprepare,APL_reservoirengineering,mi2024stable}. Notably, Ref. \cite{mi2024stable} realized state preparation in a 35-spin TFIM with Trotterization. This suggests that our protocol may be testable and applicable on existing platforms.
}

{To connect our theoretical protocol more directly with experiments, we provide in \cite{supplemental_material} proposals that relate $r$ to experimentally tunable parameters, offering realistic paths towards implementation.}

\textit{Concluding remarks.}---We have introduced a general framework
for accelerating relaxation in open quantum systems via temporary reset, applicable to arbitrary initial states. 
The proposed protocol can suppress multiple relaxation modes simultaneously, enabling
enhanced control over the relaxation dynamics. {As a practical example, we demonstrated that the relaxation of a TFIM can be significantly accelerated by using a simple depolarizing channel. This example highlights a powerful feature of our protocol: leveraging easily prepared states to accelerate the preparation of states that are difficult to reach. Furthermore, introducing temporary dephasing noise---a different type of channel---can also accelerate relaxation (End Matter). This suggests the broader applicability of our central idea: temporarily coupling the system to various quantum channels may provide a general route to enhanced relaxation.} In future studies,
our approach could be extended to certain non-Markovian and Floquet dissipative systems (e.g., using ideas in \cite{21PRL_Qmpemba,NonMarkov2025}). 
Overall, our results establish {temporary} reset as a powerful and experimentally
feasible tool for controlling relaxation timescales in open quantum
dynamics.

We are grateful to anonymous referees for their valuable comments, which greatly enhance the quality of this work. This work is partly supported by the Innovation Program for Quantum
Science and Technology (2021ZD0303306), MOST(2022YFA1303100).

\appendix

\section*{End Matter}

\textit{Derivation of the modified Lindbladian $\mathcal{L}_r$.}---{Here, we derive Eq. \eqref{modified_L} of the main text.
Let $\{|\phi_i\rangle\}_{i=1}^{d}$ be an orthonormal basis, satisfying $\sum_i |\phi_i\rangle\!\langle\phi_i|=\mathbb I$. The reset state is generally written as $\rho_{\delta}=\sum_{\alpha}p_\alpha
|\psi_\alpha\rangle\!\langle\psi_\alpha|$. 
Introduce jump operators
\begin{equation*}
  J_{i,\alpha}^{r}=\sqrt{r\,p_\alpha}|\psi_\alpha\rangle\!\langle\phi_i|, 
  \quad i=1,\dots ,d .
  \label{eq:J_pure}
\end{equation*}
The associated Lindblad term is
\begin{equation}
  \mathcal D_r(\rho)=\sum_{i,\alpha}\!\bigl(J_{i,\alpha}^{r}\rho\,J_{i,\alpha}^{r\dagger}
  -\tfrac12\{J_{i,\alpha}^{r\dagger}J_{i,\alpha}^{r},\rho\}\bigr).
  \label{eq:D_pure_def}
\end{equation}
Direct evaluation gives
\begin{align*}
  &\sum_{i,\alpha}J_{i,\alpha}^{r}\rho\,J_{i,\alpha}^{r\dagger}
  =r\,\rho_\delta\,\text{Tr}[\rho]\\
 &-\tfrac12\sum_{i,\alpha}\{J_{i,\alpha}^{r\dagger}
  J_{i,\alpha}^{r},\rho\}=-r\,\rho,
\end{align*}
where $\text{Tr}[\rho]=\sum_i\langle\phi_i|\rho|\phi_i\rangle$, $\sum_i |\phi_i\rangle\!\langle\phi_i|=\mathbb I$ and $\langle \psi_{\alpha}|\psi_{\alpha}\rangle=1$ have been used. This yields $\mathcal D_r(\rho)=
  r\bigl[\text{Tr}[\rho]\,\rho_\delta-\rho\bigr]$.
Hence, we arrive at Eq. \eqref{modified_L},
\begin{equation}
  \mathcal L_r(\rho)=
  \mathcal L(\rho)+r\text{Tr}[\rho]\,\rho_\delta-r\rho.
\end{equation}
}

{\textit{Derivation of the modified coefficients.}---Here, we derive Eq. \eqref{modified_coeff}, the explicit form of the modified coefficients with reset.} Acting the modified semi-group $e^{t\mathcal{L}_{r}}$ on both sides of
\begin{equation}
\rho_{0}=\rho_{\rm ss}+\sum_{k=2}^{d^{2}}\text{Tr}(L_{k}^{\dagger}\rho_{0})R_{k}
\end{equation}
yields 
\begin{equation}\label{rho_reset}
\rho^{r}(t)  = e^{t\mathcal{L}_{r}}\rho_{0}= e^{t\mathcal{L}_{r}}\rho_{\rm ss}+\sum_{k=2}^{d^{2}}\text{Tr}(L_{k}^{\dagger}\rho_{0})e^{(\lambda_{k}-r)t}R_{k}.
\end{equation}
To proceed, we use an explicit form of the steady
state under reset dynamics \cite{18PRE_spectral}
\begin{equation}
\rho_{\rm ss}^{r}=\lim_{t\rightarrow\infty}\rho^{r}(t)=\rho_{\rm ss}+\sum_{k=2}^{d^{2}}\frac{rd_k}{r-\lambda_{k}}R_{k}.\label{reset_state}
\end{equation}
Then, expressing the steady-state condition $e^{t\mathcal{L}_{r}}\rho_{\rm ss}^{r}=\rho_{\rm ss}^{r}$ with Eq. \eqref{reset_state} and applying Eq. \eqref{spectral_shift}, we get
\begin{equation*}
e^{t\mathcal{L}_{r}}\rho_{\rm ss}+\sum_{k=2}^{d^{2}}\frac{rd_k}{r-\lambda_{k}}e^{(\lambda_{k}-r)t}R_{k} 
=\rho_{\rm ss}+\sum_{k=2}^{d^{2}}\frac{rd_k}{r-\lambda_{k}}R_{k},
\end{equation*}
from which we solve $e^{t\mathcal{L}_r}\rho_{\rm ss}$. Substituting $e^{t\mathcal{L}_r}\rho_{\rm ss}$ into Eq. \eqref{rho_reset} and comparing with Eq. \eqref{spectral_modified}, we identify modified coefficients $c_{k}^r(t)$ defined in Eq. \eqref{modified_coeff}.  
 
{\textit{Realizing the reset protocol with arbitrary state preparation processes.}---We now show that any state preparation protocol could approximately realize the reset channel in our scheme via a Trotterization-based construction. Notably, Trotterization has already been implemented experimentally in superconducting quantum circuits \cite{21PRL_Trottersimu}, IBM Quantum's hardware \cite{22PRXQuantum_Trottersimu} and even quantum many-body platform \cite{mi2024stable} to generate dissipative dynamics, such as the dissipative TFIM. Moreover, tensor network methods offer promising avenues for extending Trotterization to more complex open quantum systems \cite{16PRL_Trotter}. We first consider the case where the original dynamics is absent [$\mathcal{L}(\rho)=0$] to gain insight. In this case, the generator corresponding to the reset channel simply reads
\begin{equation}
    \dot{\rho}(t)=\mathcal{R}[\rho(t)]=r[\rho_{\delta}-\rho(t)].
\end{equation}
Thus, a direct integration shows that applying the reset channel for a time $t_s$ maps an initial state $\rho_0$ to a final state $\rho(t_s)$ according to
\begin{equation}\label{prep_channel}
    \rho(t_s)=e^{\mathcal{R}t_s}[\rho_0]=p\rho_0+(1-p)\rho_{\delta},\quad p:=e^{-r t_s}.
\end{equation}
That is, a reset channel with rate $r=-\ln p/t_s$ is effectively equivalent to a state preparation protocol for $\rho_{\delta}$, with
success probability $1-p$. $p$ is a classical probability generated beforehand. One simply flips a biased coin (or, equivalently, performs any local measurement that yields the desired classical randomness): with probability $1-p$ the system is driven into $\rho_{\delta}$ by any suitable operations and with the probability $p$ it is left untouched. Since there are no restrictions on the preparation method, once $\rho_{\delta}$ is locally preparable (e.g. a separable state), this operation can be implemented purely with local operations, even for many-body systems. Specifically, when $\rho_{\delta}=\mathbb{I}/d$, Eq. \eqref{prep_channel} is a depolarizing channel, which can be directly realized by a unitary 2-design \cite{supplemental_material,Mele2024introductiontohaar} without flipping a coin. 

When $\mathcal{L}\neq0$, we apply the Lie-Trotter formula \cite{lindblad1976generators}
\begin{equation}
    e^{(\mathcal{L}+\mathcal{R})t}=\lim_{n\to\infty}(e^{\mathcal{R}\frac{t}{n}}e^{\mathcal{L}\frac{t}{n}})^n.
\end{equation}
to realize the protocol. With this formula, the reset protocol during $[0,t_s]$ can be approximated as 
\begin{equation}\label{trotter_approx}
     e^{(\mathcal{L}+\mathcal{R})t_s}[\rho_0]\approx(e^{\mathcal{R}\delta t}e^{\mathcal{L}\delta t})^n[\rho_0],
\end{equation}
where $\delta t:=t_s/n\ll 1$. Using Eq. \eqref{prep_channel}, we have 
\begin{equation}
    e^{\mathcal{R}\delta t}[\rho]\approx (1-p_s)\rho + p_s \rho_{\delta},
\end{equation}
which can be interpreted as a state preparation mapping from $\rho$ to $\rho_{\delta}$, with success probability $p_s:=r\delta t$. Practically, $r$ is determined by $p_s$ and $\delta t$, whose values are set by experimentalists.

Thus, the reset protocol may be achieved experimentally by performing the aforementioned state preparation operations stroboscopically: at each discrete time point $t_i=i\delta t,\ (i=1,...,n)$ within $[0,t_s]$, one applies the state preparation mapping from $\rho (t_i)$ to $\rho_{\delta}$ with a given probability $p_s$. There are no constraints on the preparation time $\tau_{\rm prep}$ for $\rho_{\delta}$, but on average it adds a cost of $np_s\tau_{\rm prep}$ to the tailored relaxation timescale. Therefore, it is desirable to choose a $\rho_{\delta}$ that can be prepared efficiently, e.g. the maximally mixed state in our TFIM example. This state is convenient to prepare: coupling to an infinite-temperature bath can achieve it in $\mathcal{O}(\ln N)$ time \cite{24arxiv_optimal} with minimal resources in a $N$-body system---typically negligible compared to the system’s intrinsic relaxation time. Local control methods could in principle realize it in constant time, though with resources scaling with $N$.

This Trotterization-based method admits a natural interpretation at the level of stochastic trajectories: for a Poissonian reset process with rate $r$, the system is stochastically reset to the target state with probability $r\delta t$, in each small interval $\delta t$. 

The approximation error 
is of order $\mathcal{O}(t_s^2/n)$ \cite{85PRB_suzuki,lloyd96science,21PRX_Trotter}. Explicitly, we establish a rigorous upper bound \cite{supplemental_material}:
\begin{equation}
    ||e^{(\mathcal{L}+\mathcal{R})t_s}-(e^{\mathcal{R}\delta t}e^{\mathcal{L}\delta t})^n||\leq\frac{t_s^2}{2n}||[\mathcal{L,R}]||,
\end{equation}
where $||\cdot||$ denotes any norm that is contractive under Lindbladian evolution, such as the trace norm or the diamond norm. This bound holds for arbitrary Lindbladian $\mathcal{L}$ and $\mathcal{R}$. In our specific case, {
\begin{equation}
    [\mathcal{L,R}]\rho:=[\mathcal{LR}-\mathcal{RL}]\rho=r{\rm Tr}(\rho)\mathcal{L}(\rho_{\delta}),
\end{equation}
where ${\rm Tr}(\mathcal{L}(\rho))=0$ is used.} Hence, a large $n$ is not required when $t_s\ll 1$, $rt_s\ll 1$ or $\rho_{\delta}$ is close to $\rho_{\rm ss}$. A small $t_s$ with moderate $r$ offers a practical regime that may yield substantial acceleration and simultaneously reduce the required number of Trotter steps.

Additionally, higher-order approximations could be used to reduce the error for fixed $n$. For instance, the second-order Suzuki-Trotter formula $e^{(\mathcal{L}+\mathcal{R})t_s}\approx(e^{\mathcal{L}\delta t/2}e^{\mathcal{R}\delta t}e^{\mathcal{L}\delta t/2})^n$ suppresses the error to $\mathcal{O}(t_s^3/n^2)$ \cite{85PRB_suzuki,21PRX_Trotter}.


\textit{Dephasing noise can accelerate the relaxation of the TFIM.---}Here, we add dephasing noise along a single axis (typically the $z$-axis) to the TFIM. The corresponding jump operators are $L^{(\phi)}_i = \sqrt{\gamma_{\phi}}\sigma_i^z, \quad i = 1,\dots,N$.
This contributes to the total Lindbladian via an extra term
\begin{equation}
\mathcal{L}_{\phi}[\rho] = \gamma_{\phi}\sum_{i=1}^N \left( \sigma_i^z \rho \sigma_i^z - \rho \right),
\end{equation}
which could be interpreted as a partial and local reset channel (only the $z$-axis is affected, and only local one-body jump operators are involved). As shown in Fig. \ref{3}, the relaxation is accelerated significantly.
\begin{figure}
    \centering
    \includegraphics[width=1.0\linewidth]{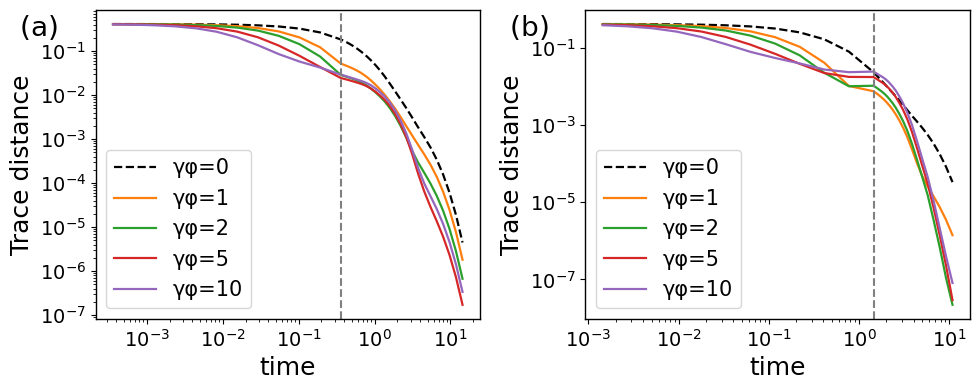}
    \caption{Dephasing noise induces acceleration of relaxation in a 5-site TFIM. Parameters: (a) $J=1.0, \ g=2.0,\ \gamma =0.5,\ \beta=0.1J,\ t_s=0.2\tau_2$ (b) $J=1.0, \ g=2.0,\ \gamma =0.5,\ \beta=0.1J,\ t_s=0.8\tau_2$.}
    \label{3}
\end{figure}

\bibliography{bibfile}

\end{document}


\begin{CJK*}{UTF8}{gbsn}
\title{Supplemental Material for ``Accelerating Quantum Relaxation via Temporary Reset: A Mpemba-Inspired Approach''
}

\author{Ruicheng Bao}
\email{Contact author: ruicheng@g.ecc.u-tokyo.ac.jp}
\affiliation{Department of Chemical Physics \& Hefei National Laboratory, University
of Science and Technology of China, Hefei 230088, China}
\affiliation{Department of Physics, Graduate School of Science, 
The University of Tokyo, Hongo, Bunkyo-ku, Tokyo 113-0033, Japan}

\author{Zhonghuai Hou}
\email{Contact author: hzhlj@ustc.edu.cn}
\affiliation{Department of Chemical Physics \& Hefei National Laboratory, University
of Science and Technology of China, Hefei 230088, China}
\maketitle
\end{CJK*}

This Supplementary Material includes the evolution equation for observables under reset, derivation of the sufficient condition for reset-induced Mpemba effect, a complementary analysis of the two-state model, an implementation proposal, error bounds of the Trotterization, and additional numerical results. 
\vspace{12pt}

\section{Accelerating relaxation of quantum observables} In experimental
settings, observables rather than full density matrices are typically
measured. Our protocol naturally extends to controlling the relaxation
dynamics of arbitrary quantum observables. The time evolution of an
observable $O$ under unperturbed dynamics reads:

\begin{equation}
\langle O\rangle(t)={\rm Tr}[Oe^{\mathcal{L}t}(\rho_{0})]=\langle O\rangle_{ss}+\sum_{k=2}^{d^{2}}c_{k}e^{\lambda_{k}t}{\rm Tr}[OR_{k}].
\end{equation}
Applying our reset protocol, the modified observable dynamics becomes

\begin{align}
\langle O\rangle^{r}(t) & =\langle O\rangle_{ss}+\sum_{k}c_{k}^{r}(t)e^{\lambda_{k}t}\Theta(t_{s}-t){\rm Tr}[OR_{k}]\nonumber \\
 & +\sum_{k}c_{k}^{r}(t_{s})e^{\lambda_{k}t}\Theta(t-t_{s}){\rm Tr}[OR_{k}].
\end{align}
Thus, the same reset condition $c_{2}^{r}(t_{s})\approx0$ also ensures
fast convergence of $\langle O\rangle(t)$ to its steady-state value
$\langle O\rangle_{ss}$. This allows for direct experimental application
of our protocol at the level of observables, without requiring full
state tomography.
{
\section{Derivation of the sufficient conditions ${\rm Re}(c_k^*d_k)< |c_k|^2$}
Recall that the definition of the modified coefficient is
\begin{equation}\label{coefficient_sm}
    c_{2}^{r}(t)=\left[c_{2}-\frac{r\cdot d_{2}}{r-\lambda_{2}}\right]e^{-rt}+\frac{r\cdot d_{2}}{r-\lambda_{2}}e^{-\lambda_{2}t},
\end{equation}
which can be a complex number in general. We want to derive the condition such that
\begin{equation}\label{mpemba_condition}
    g(t_s)=|c_2^r(t_s)|-|c_2|<0
\end{equation}
holds for any $t_s\in(0, t_c]$. Here, $t_c>0$ is a threshold value that make $|c_2(t_c)|=|c_2|$. If this condition is satisfied, the modified relaxation dynamics will be equivalent to the dynamics governed by a smaller $|c_2|$ compared to the original dynamics. That is, the relaxation is accelerated by the temporary reset, in the spirit of the Mpemba effect.

Notably, $g(0)=0$, which implies that the sufficient condition for Eq. \eqref{mpemba_condition} to hold for any $t_s\in(0, t_c]$, i.e., all $t_s$ that are not too large, is given by 
\begin{equation}
    \left.\frac{dg}{dt}\right|_{t=0}<0.
\end{equation}
Since $|c_2|$ is a constant, the condition is equivalent to 
\begin{equation}\label{sufficient_con}
    \left.\frac{d[|c_2^r(t)|^2]}{dt}\right|_{t=0}<0.
\end{equation}
Using the chain rule, we have
\begin{equation}
    \frac{d[|c_2^r(t)|^2]}{dt}=\frac{d[c_2^rc_2^{r*}]}{dt}=(c^{r}_2)'c_2^{r*}+[(c^{r}_2)'c_2^{r*}]^*=2{\rm Re}[(c^{r}_2)'c_2^{r*}],
\end{equation}
so we only have to calculate $(c^{r}_2)':=(dc_2^r/dt)|_{t=0}$. A straightforward calculation using Eq. \eqref{coefficient_sm} yields
\begin{equation}
    \left. \frac{dc_2^r}{dt}\right|_{t=0}=-r(c_2-d_2), \quad c_2^{r*}(0)=c_2^*.
\end{equation}
Therefore, the sufficient condition Eq. \eqref{sufficient_con} is equivalent to
\begin{equation}
    \left.\frac{d[|c_2^r(t)|^2]}{dt}\right|_{t=0}=2{\rm Re}[(c^{r}_2)'c_2^{r*}]|_{t=0}=-2r{\rm Re}[|c_2|^2-c_2^*d_2]< 0.
\end{equation}
Because $r\geq 0$, a rearrangement leads to the desired condition 
\begin{equation}
    {\rm Re}(c_2^*d_2)<|c_2|^2
\end{equation}
in the main text. It is straightforward to see that
\begin{equation}
    {\rm Re}(c_k^*d_k)<|c_k|^2
\end{equation}
is the sufficient condition for the existence of another threshold value $t_{c,k}>0$ such that 
\begin{equation}
    |c_k^r(t_s)|-|c_k|<0
\end{equation}
holds for any $t_s\in(0, t_{c,k}]$, given that $c_k^r(t_s)$ has the same structure for any $k\geq2$. 

If the conditions hold for multiple modes $k\in S_m:={2,\dots,m}$, then any $t_s\in(0,t_c^*]$ will suppress all of them simultaneously, where $t_c^*\equiv \min_{k\in S_m} t_{c,k}$. In Section VII B, we numerically show that multiple such conditions for different $k$ can typically be satisfied simultaneously.

}

\section{Spectral analysis of the two-state model} The Lindblad master equation for the two-state model can be rewritten as a matrix equation
for the vector $\vec{\rho}(t)=[\rho_{00}(t),\rho_{01}(t),\rho_{10}(t),\rho_{11}(t)]^{T}$:
\begin{equation}
\frac{d\vec{\rho}(t)}{dt}=L\vec{\rho}(t),
\end{equation}
where $L$ is a $4\times4$ Liouvillian matrix in the Fock-Liouvillian
space, taking the form 
\begin{equation}
L=\left(\begin{array}{cccc}
-\gamma_{0} & -i\Omega & i\Omega & \gamma_{1}\\
-i\Omega & -iE-\frac{\gamma_{0}+\gamma_{1}}{2} & 0 & i\Omega\\
i\Omega & 0 & iE-\frac{\gamma_{0}+\gamma_{1}}{2} & -i\Omega\\
\gamma_{0} & i\Omega & -i\Omega & -\gamma_{1}
\end{array}\right).
\end{equation}
In the simple case $\Omega=0$, the right eigenvalues and eigenvectors
of the matrix $L$ are given by 
\begin{align}
\lambda_{1} & =0,\ \vec{R}_{1}=(\gamma_{1},0,0,\gamma_{0})^{\text{T}}\nonumber \\
\lambda_{4} & =-\gamma_{0}-\gamma_{1},\ \vec{R}_{4}=(-1,0,0,1)^{\text{T}}\nonumber \\
\lambda_{2} & =\frac{1}{2}(-2i-\gamma_{0}-\gamma_{1}),\ \vec{R}_{2}=(0,1,0,0)^{\text{T}}\nonumber \\
\lambda_{3} & =\frac{1}{2}(2i-\gamma_{0}-\gamma_{1}),\ \vec{R}_{3}=(0,0,1,0)^{\text{T}}
\end{align}
Likewise, the left eigenvalues and eigenvectors are given by 
\begin{align}
\lambda_{1}^{\star} & =0,\ \vec{L}_{1}=(1,0,0,1)^{\text{T}}\nonumber \\
\lambda_{4}^{\star} & =-\gamma_{0}-\gamma_{1},\ \vec{L}_{4}=(-\gamma_{0},0,0,\gamma_{1})^{\text{T}}\nonumber \\
\lambda_{2}^{\star} & =\frac{1}{2}(2i-\gamma_{0}-\gamma_{1}),\ \vec{L}_{2}=(0,1,0,0)^{\text{T}}\nonumber \\
\lambda_{3}^{\star} & =\frac{1}{2}(-2i-\gamma_{0}-\gamma_{1}),\ \vec{L}_{3}=(0,0,1,0)^{\text{T}}.
\end{align}
{Notably, the eigenvectors proposed here have not yet been normalized. They should be normalized to satisfy the biorthogonal condition ${\rm tr}(L_k^{\dagger}R_h)=\delta_{kl}$.
Setting $c^r_4(t_s)=0$ for $t_{s}$ yields its expression:
 \begin{equation}
t_{s}=\frac{\ln\left[1+\frac{\gamma_{1}k_{\text{tot}}}{r}\left(\frac{1}{\gamma_{0}+\gamma_{1}e^{(\beta_{0}-\beta_{\text{env}})E}}-\frac{1}{\gamma_{0}+\gamma_{1}}\right)\right]}{k_{\text{tot}}},
\end{equation}
from which one can see that the critical time would not be affected
by the coherence terms of the initial state. Here, the total transition rate is $k_{\text{tot}}=r+\gamma_{0}+\gamma_{1}$. The critical time is used in the Example 1 of the main text.

\section{Elimination of relaxation modes via fine-tuning}
Here, we show that our reset protocol can fully eliminate the dominant relaxation modes, if the fine-tuning of control parameters are allowed and the initial state and the dynamics of the system of interest are known \textit{a priori}, as in previous studies.
First, recall that the characteristic relaxation timescale is given by $\tau_{{\rm {\rm rel}}}\sim\frac{1}{|\text{Re}(\lambda_{2})|}$ because in the long-time limit, the relaxation is dominated
by the slowest decaying mode, and the deviation from stationarity
is $||\rho(t)-\rho_{\rm ss}||\sim\exp\left(-|\text{Re}\lambda_{2}|t\right)$.

By choosing an optimal $t^*_{s}$, we can set $c_{2}^{r}(t^*_{s})=0$, thereby eliminating
the overlap with the slowest decaying mode. Solving the equation
\begin{equation}
    c_{2}^{r}(t)\equiv\left[c_{2}-\frac{r\cdot d_{2}}{r-\lambda_{2}}\right]e^{-rt}+\frac{r\cdot d_{2}}{r-\lambda_{2}}e^{-\lambda_{2}t}=0
\end{equation}
yields
\begin{equation}
t^*_{s}(r)=\frac{1}{r-\lambda_{2}}\ln\left[1-\frac{c_{2}(r-\lambda_{2})}{r\cdot d_{2}}\right].\label{ts}
\end{equation}
By an appropriate choice of $\rho_{\delta}$ such that $c_2/d_2\leq0$, $t^*_{s}(r)\geq 0$ can be assured. The protocol with an optimal $t_s=t^*_s$ reduces
the relaxation time to 
\begin{equation}
\tau_{{\rm rel}}\sim t^*_s+ \frac{1}{|{\rm Re}(\lambda_{3})|},
\end{equation}
which is generically shorter than the unperturbed timescale $\tau_{{\rm {\rm rel}}}\sim\frac{1}{|\text{Re}(\lambda_{2})|}$.

For cases when the dominant modes form a complex-conjugate pair, eliminating the entire complex mode reduces to a single
complex condition:
\begin{equation}
c_{2}^{r}(t_{s})=0\Leftrightarrow\text{Re}[c_{2}^{r}(t_{s})]=\text{Im}[c_{2}^{r}(t_{s})]=0,\label{condition}
\end{equation}
because $c_{2}^{r}(t_{s})=0\Rightarrow c_{2}^{r,*}(t_{s})=[c_{2}^{r}(t_{s})]^{*}=0$.
{In some cases,} this condition { can be exactly satisfied} by tuning $r$ and $t_{s}$. {Although Eq. \eqref{condition} does not always admit an exact solution,} in practice, it suffices to reduce $|c_{2}^{r}(t_{s})|$ to an exponentially
small value ($\text{Re}[c_{2}^{r}(t_{s})]\ll1,\ \text{Im}[c_{2}^{r}(t_{s})]\ll1$),
which still leads to exponential acceleration of relaxation. Notably, a similar analysis was provided in \cite{Busiello_2021} for classical systems and for specific cases where the detailed balance is satisfied, which ensures all eigenvalues are real.

{
\section{Experimental implementation proposals}
\subsection{Multilevel, few-body systems: reservoir engineering}
The required interaction Hamiltonian between the system of interest and the ancilla qubits system is given by 
\begin{equation}
    H_{\rm int}=\sum_{i=1}^M g_i(J_i^{\dagger}\otimes\sigma_i^{-}+\rm h.c.),
\end{equation}
where $J_i^{\dagger}$ is the reset jump operator and the coupling strength $g_i$ between the system of interest and the $i$-th ancilla qubit can be tuned experimentally, e.g. using the superconducting quantum interference device (SQUID) \cite{14prl_SQUID}. {Here, $J_i^{\dagger}$ acts on the Hilbert space of the system, while $\sigma_i^{-}$ acts on the Hilbert space of the $i$-th ancilla qubit. Identity operators acting on all other subsystems are implicit. In this sense, each term in Eq.\,(22) can be regarded as the local interaction Hamiltonian between the system and the $i$-th ancilla qubit, and the total interaction Hamiltonian is their sum, $H_{\mathrm{int}}=\sum_i H_{\mathrm{int}}^{(i)}$. Since the jump operators $J_i$ and couplings $g_i$ may differ with $i$, the contributions from different ancilla qubits are generally distinct.}

If each of the $M$ qubits independently dissipates according to the local Lindbladian $\mathcal{L}(\rho)=\tau_{a}^{-1}(\sigma_i^{-}\rho\sigma_i^{+}-\{\sigma_i^{+}\sigma_i^-,\rho\}/2)$ with a fast timescale $\tau_a\ll 1$, then the reset rate would be \cite{16pra_universalsimulation}
\begin{equation}
    r=4g^2\tau_a,
\end{equation}
where we set equal couplings $g_i=g$.
Thus, the reset rate $r$ can be tuned experimentally by varying $g$ and the relaxation timescale $\tau_a$ of the ancilla.

Take our two-state system (Example 1 of the main text) as an example.
Consider an arbitrary pure state $|\delta\rangle=\cos(\tfrac{\theta}{2})|0\rangle+e^{i\varphi}\sin(\tfrac{\theta}{2})|1\rangle$ as the reset target $\rho_{\delta}$, there are only two jump operators involved:
\begin{equation}
  J_0=\sqrt{r}\,|\delta\rangle\!\langle 0|\,,\qquad
  J_1=\sqrt{r}\,|\delta\rangle\!\langle 1|\,.
  \label{eq:reset_J_pure}
\end{equation}
{Notably, $\theta$ and $\varphi$ are tunable parameters. In practical situations, the reset state $|\delta\rangle$ is typically chosen so that the jump operators do not involve transitions between non-orthogonal states, which are challenging to realize experimentally. For instance, in the present case, we could choose $\theta =0$ so that $|\delta\rangle=|0\rangle$, as in the main text.} To realize the jump operators, we only need to introduce two ancilla qubits  $A_0$ and $A_1$ with ground and excited states $|g\rangle,|e\rangle$ and fast
spontaneous relaxation $\ket{e}\!\to\!\ket{g}$ at rate $\kappa=1/\tau_a\gg1$. Then, 
one should realize the coherent couplings,
\begin{equation}
  H_{\rm int}=\sum_{i=0}^{1} \left[g_i\,\big(J_i^\dagger\otimes\sigma_i^- \big)+ g_i\,\big(J_i\otimes\sigma_i^+\big)\right],
  \label{eq:Hint_qubit}
\end{equation}
where $\sigma_i^-=|g\rangle\!\langle e|$ acts on ancilla $A_i$, and $J_i$ are the system jump operators from
Eq.~\eqref{eq:reset_J_pure}. Then, the effective reset rate reads $r\approx4g^2\tau_a$. 

Here, the interaction strength $g$ can be tuned via microwave or laser amplitude control in Rydberg atoms \cite{sevinccli2014microwave,morgado2021quantum}; SQUID-based tunable couplers in superconducting circuits.

When $\rho_{\delta}=$ is a mixed state written as $\rho_\delta=\sum_{\alpha=1}^2 p_\alpha|\psi_\alpha\rangle\!\langle\psi_\alpha|$ (rank~$\le 2$ for a qubit), the protocol may be realized by introducing 4 ancilla qubits; Interestingly, it may also be realized via time-multiplex the pure state case using the same two ancillas: for a fraction $p_1$
of the reset window, program $J_i=\sqrt{r}\,|\psi_1\rangle\!\langle i|$, and for the remaining fraction $p_2=1-p_1$,
program $J_i=\sqrt{r}\,|\psi_2\rangle\!\langle i|$ (with identical $g$ so the rate is $r$ in both sub-windows).
If the switching period is short compared to the system relaxation time, the effective generator equals
the convex combination
\begin{equation}
    p_1\,\mathcal{D}_{|\psi_1\rangle} + p_2\,\mathcal{D}_{|\psi_2\rangle}
= r\!\left[\mathrm{Tr}[\rho]\rho_\delta-\rho\right]
\end{equation}
thereby realizing the desired mixed-state reset still with only two ancilla qubits. Here, $\mathcal{D}_{|\psi_i\rangle}=r\!\left[\mathrm{Tr}[\rho]|\psi_1\rangle\langle\psi_i|-\rho\right]$.

\subsection{Many-body systems:}
In the End Matter, we have described in details how to connect the reset rate $r$ with tunable parameters through Trotterization. Here, we discuss the realization of a specific case to better illustrate the protocol, where $\rho_{\delta}=\mathbb{I}/d$ as in Example 2 of the main text. In this case, the reset channel becomes the depolarizing channel. There are two primary methods to realize this channel, which we detail below.

The first approach, often used for noise characterization, realizes the depolarizing channel through a twirling operation \cite{Mele2024introductiontohaar}:
\begin{equation}
    \mathbb{E}_{U\sim \nu}[U^{\dagger} \Phi(U\rho U^{\dagger})U]=p_{\Phi}\rho+(1-p_{\Phi}){\rm Tr}(\rho)\frac{\mathbb{I}}{d},
\end{equation}
where $p_{\Phi}=\frac{d^2F_{\Phi}-1}{d^2-1}\in [0,1]$. $F_{\Phi}:=\frac{1}{d^2}\langle \Omega|\Phi \otimes \mathcal{I}(|\Omega\rangle\langle \Omega|)|\Omega\rangle$ is the entanglement fidelity of the quantum channel $\Phi$ and $U\sim \nu$ implies that the unitary $U$ is randomly generated from an (approximate) 2-design \cite{09PRA_2design}. Here, $|\Omega\rangle$ is the maximally entangled state. The quantity $1-p_{\Phi}$ plays the role of the success probability in our protocol, and can be tuned by choosing different quantum channel $\Phi$. This operation is routine in quantum information science, especially in Randomized Benchmarking \cite{Mele2024introductiontohaar}. Notably, the random unitary $U$ here may be achieved within a timescale irrelevant to the system size \cite{foxman2025random}.

A second, more direct method for simulation or engineered noise implements the depolarizing channel via the stochastic application of Pauli operators. This approach leverages classical randomness to construct the channel's effect, which is precisely the general method for implementing the reset channel discussed in the End Matter. For a many-body system, this is typically implemented by applying noise to each qubit independently. Concretely, for each qubit, the identity operation $\mathcal{I}$ is applied with probability $1-q$, and one of the Pauli operators ($\sigma_{x,y,z}$) is applied with probability $q$.

Our reset protocol can be realized by periodically applying the depolarizing channel with a period of $\delta t$, using either of the two methods described above. This Trotterization technique connects the reset rate $r$ to the experimentally tunable parameters via the relation $r=p/\delta t$, where the success probability $p$ is the probability of a non-identity operation per step (i.e., $p=1-p_{\Phi}$ for the twirling method, and $p=q$ for the stochastic method). Such Trotterization has been experimentally performed in complex many-body quantum systems \cite{mi2024stable}.

For the stochastic implementation, a single Trotterized evolution may be sufficient to realize the desired dynamics. An evolution over a total time $t_s$ is decomposed into a large number of discrete steps, $N=t_s/\delta t$. At each step, an independent random Pauli error is potentially applied based on a classical probability. By the law of large numbers, this time-series of stochastic operations within a single trajectory accurately approximates the ensemble-averaged effect of the ideal depolarizing channel when $N\gg1$.
}

\section{Proof of the Trotter error bound presented in End Matter}
Following \cite{lloyd96science}, we can first provide an intuitive analysis of the error by considering the Taylor expansion:
\begin{equation}
    e^{(\mathcal{L}+\mathcal{R})t_s}=(e^{\mathcal{R}\delta t}e^{\mathcal{L}\delta t})^n+\frac{t_s^2}{2n}[\mathcal{L,R}]+\mathcal{O}\left(\frac{t_s^3}{n^2}\right),
\end{equation}
which roughly shows that the leading error is of the order of $\mathcal{O}(t_s^2/n)$, as mentioned in the End Matter. However, it is unclear whether high-order terms should be taken into account in general, particularly in some large quantum systems with complex interactions.

In what follows, we prove the statement in End Matter that the error of the approximation from Lie-Trotter formula is upper bounded as
\begin{equation}\label{error_bound}
    ||e^{(\mathcal{L}+\mathcal{R})t_s}-(e^{\mathcal{R}\delta t}e^{\mathcal{L}\delta t})^n||\leq\frac{t_s^2}{2n}||[\mathcal{L,R}]||,
\end{equation}
where $||\cdot||$ can be the trace norm or the diamond norm. 

\textit{Proof of the error bound.---}For two arbitrary bounded operators $\mathcal{L}$ and $\mathcal{R}$, we have \cite{85PRB_suzuki}
\begin{equation}\label{suzuki}
e^{x(\mathcal{L}+\mathcal{R})} - e^{x\mathcal{L}} e^{x\mathcal{R}}
= \int_0^x dt \int_0^t ds\, e^{t\mathcal{L}} e^{(t-s)\mathcal{R}} [\mathcal{R}, \mathcal{L}]\, e^{s\mathcal{R}} e^{(x-t)(\mathcal{L}+\mathcal{R})},
\end{equation}
which follows from Kubo's identity 
\begin{equation}
    [\mathcal{L},e^{t\mathcal{R}}]=\int_{0}^te^{(t-s)\mathcal{R}}[\mathcal{L},\mathcal{R}]e^{s\mathcal{R}}ds.
\end{equation}
Taking the trace norm or diamond norm on both sides of Eq. \eqref{suzuki}, applying the triangle inequality first, and then using sub-multiplicativity of matrix norm ($||AB||\leq ||A||\cdot ||B||$) yields
\begin{equation}\label{bound1}
    e^{x(\mathcal{L}+\mathcal{R})} - e^{x\mathcal{L}} e^{x\mathcal{R}}\leq \int_0^x dt \int_0^t ds\, ||e^{t\mathcal{L}}||\cdot ||e^{(t-s)\mathcal{R}}||\cdot ||[\mathcal{R}, \mathcal{L}]||\cdot ||e^{s\mathcal{R}}||\cdot ||e^{(x-t)(\mathcal{L}+\mathcal{R})}||.
\end{equation}
To proceed, we employ the contractivity of the trace norm or diamond norm under any Lindbladian $\mathcal{K}$, i.e.,
\begin{equation}\label{contractivity}
    ||e^{t\mathcal{K}}||\leq 1,
\end{equation}
for any $t\geq0$. Notice that $t-s\geq0$, $x-t\geq 0$ due to their limits of integration. Thus, combining Eqs. \eqref{bound1} and \eqref{contractivity} and taking $x=\delta t=t_s/n$ leads to the bound for single-step error: 
\begin{equation}\label{single_bound}
    ||e^{(\mathcal{L}+\mathcal{R})\delta t} - e^{\mathcal{L}\delta t} e^{\mathcal{R}\delta t}|| \leq \frac{t_s^2}{2n^2}||[\mathcal{L},\mathcal{R}]||.
\end{equation}
Defining $g=e^{(\mathcal{L}+\mathcal{R})\delta t},\ f= e^{\mathcal{L}\delta t} e^{\mathcal{R}\delta t}$, our goal of proving the $n$-step bound Eq. \eqref{error_bound} can be rewritten as 
\begin{equation}
    ||g^n-f^n||\leq \frac{t_s^2}{2n}||[\mathcal{L},\mathcal{R}]||.
\end{equation}
Taking the norm on both sides of the telescoping identity
\begin{equation}
    g^{n}-f^n=(g-f)(g^{n-1}+g^{n-2}f+...+gf^{n-2}+f^{n-1})=(g-f)\sum_{k=0}^{n-1}g^{n-1-k}f^k,
\end{equation}
and using the triangle inequality, sub-multiplicativity and the contractivity directly yields
\begin{equation}
    ||g^n-f^n||\leq ||g-f||\sum_{k=0}^{n-1}||g^{n-1-k}||\cdot ||f^k||\leq n||g-f||.
\end{equation}
We thus complete the proof by using the single-step bound Eq. \eqref{single_bound}.

Our bounds are valid for any Lindbladian $
\mathcal{L}$ and $\mathcal{R}$. To our knowledge, this bound has not been established before. A similar previously known bound only holds for unitary dynamics \cite{21PRX_Trotter,22PRL_trotter}. In \cite{85PRB_suzuki}, an error bound for any operator is provided, which will be exponentially loose when specifically applied to two Lindbladians (because the properties of the Lindbladian are not considered there). Notably, a single-step version ($n=1$ and $\delta t=t/n=t$) of Eq. \eqref{error_bound} was proved in \cite{16PRL_Trotter}. However, only error scaling is provided there for $n$-step cases, without rigorous error bounds. 

In our specific case, $\mathcal{R}[\rho]=r[\text{Tr}(\rho)\rho_{\delta}-\rho]$, thus
\begin{equation}
    [\mathcal{L,R}]\rho:=[\mathcal{LR}-\mathcal{RL}]\rho=r{\rm Tr}(\rho)\mathcal{L}(\rho_{\delta})
\end{equation}
for the CPTP $\mathcal{L}[\rho]$.

Notably, when $\delta t\to 0$, i.e., the mapping $\rho(t)\to p_s\rho_{\delta}+(1-p_s)\rho(t)$ could occur at any time $t\in[0,t_s]$, the approximation becomes exact. Thus, the depolarizing channel used in the TFIM example could be added exactly by continuously applying an isotropic noise field.

\section{Additional numerical results}
{
\subsection{An additional example: the dissipative Dicke model}
Here we provide another example illustrating relaxation acceleration, the dissipative Dicke model, where $N$ two-level atoms collectively couple to a single lossy cavity mode. 
The full model is described by the Hamiltonian
\begin{equation}
H = \omega a^\dagger a + \Omega S_z + \frac{2g}{\sqrt{N}} (a^\dagger+a) S_x,
\end{equation}
with cavity annihilation operator $a$ and collective spin operators $S_x,S_z$, together with a single jump operator $L_c = \sqrt{\kappa}\,a.$ The Hilbert space dimension of this description is $2^N$ for the spins. Assuming that the cavity mode can be adiabatically eliminated for numerical simplicity, one obtains an effective spin-only dynamics restricted to the fully symmetric subspace with total spin $S=N/2$ (dimension $N+1$). 
The resulting master equation is
\begin{equation}
\dot\rho = -i[\tilde H,\rho] + \mathcal{D}[\tilde L_1]\rho,
\end{equation}
with effective operators
\begin{align}
\tilde H &= \Omega S_z - \frac{4\omega g^2}{4\omega^2+\kappa^2}\,\frac{S_x^2}{N},\\
\tilde L_1 &= \frac{2|g|\sqrt{\kappa}}{\sqrt{4\omega^2+\kappa^2}}\,\frac{S_x}{\sqrt{N}}.
\end{align}
This effective description reduces the dynamics to a Hilbert space of size $N+1$, which enables exact numerical simulation for large $N$. 
The same model was first employed in Ref.~\cite{21PRL_Qmpemba}, and we follow their setup; hence we do not repeat all details here. In the example, the $\rho_{\delta}$ is also chosen as the maximally mixed state $\mathbb{I}/d$ and the reset rate is $r=1.0$.

\begin{figure*}
    \centering
    \includegraphics[width=0.5\linewidth]{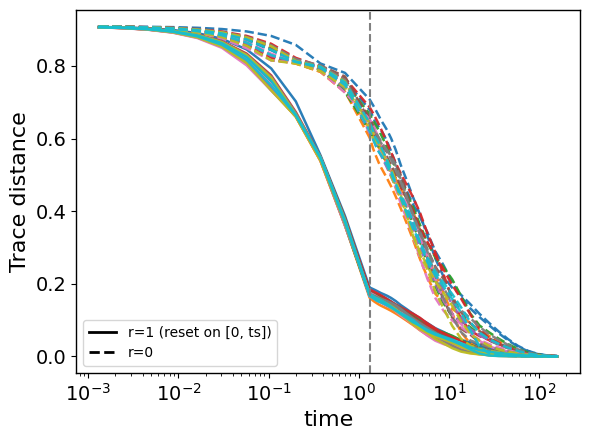}
    \caption{Acceleration of relaxation of the dissipative Dicke model with $N=10$. $20$ different initial states are randomly chosen from the Haar measure. Dashed lines: without reset; solid: with reset. Parameters: $\omega=\Omega=g=\kappa=1.0$,  $r=1.0$, $t_s=0.2\tau_2$ is denoted by the vertical dashed line,  $\rho_{\delta}=\mathbb{I}/d$.}
    \label{S_7}
\end{figure*}
}
\subsection{Additional numerical results on the dissipative transverse field Ising model}
Recall that the TFIM of length $N$ we use has the Hamiltonian
\begin{equation*}  
H =-J\sum_{i=1}^{N-1}\sigma_i^z\sigma_{i+1}^z
     -g\sum_{i=1}^{N}\sigma_i^x , 
\end{equation*}
and jump operators $J_{i,\downarrow}=\sqrt{\gamma}\,\sigma_i^-,
\quad
J_{i,\uparrow}  =\sqrt{\gamma e^{-\beta}}\;\sigma_i^+$. 
In the main text, we fix the coupling strength $J=1$, the field strength $g=1.2$,  the intrinsic rate $\gamma=0.5$, and the inverse temperature $\beta=1/k_BT=1$. Following the main text, the reset channel added here is the depolarizing channel 
\begin{equation*}
    \mathcal{R}[\rho(t)]=r\left[\frac{\mathbb{I}}{d}-\rho(t)\right],
\end{equation*}
i.e., the system state $\rho(t)$ is randomly reset to the maximally mixed state $\rho_{\delta}=\mathbb{I}/d$ with Poissonian rate $r$ when $t\in [0,t_s]$. The initial state is chosen as $\rho_0=\rho_{\rm ss}+\alpha V_2$, where $\rho_{\rm ss}$ is the steady-state density matrix, and $\alpha$ controls the distance from the initial state to the steady state.
{
\subsubsection{Accelerated relaxation dynamics with random initial states under the parameters of main text}

Using the parameters of the main text ($g=1.2,\ J=\beta=1.0,\ \gamma=0.5$),
we first test $20$ randomly chosen pure initial states and still observe significant acceleration in a $N=5$ TFIM with diferent reset rate $r$, see Fig. \ref{S_3}.

\begin{figure*}
    \centering
    \includegraphics[width=1.0\linewidth]{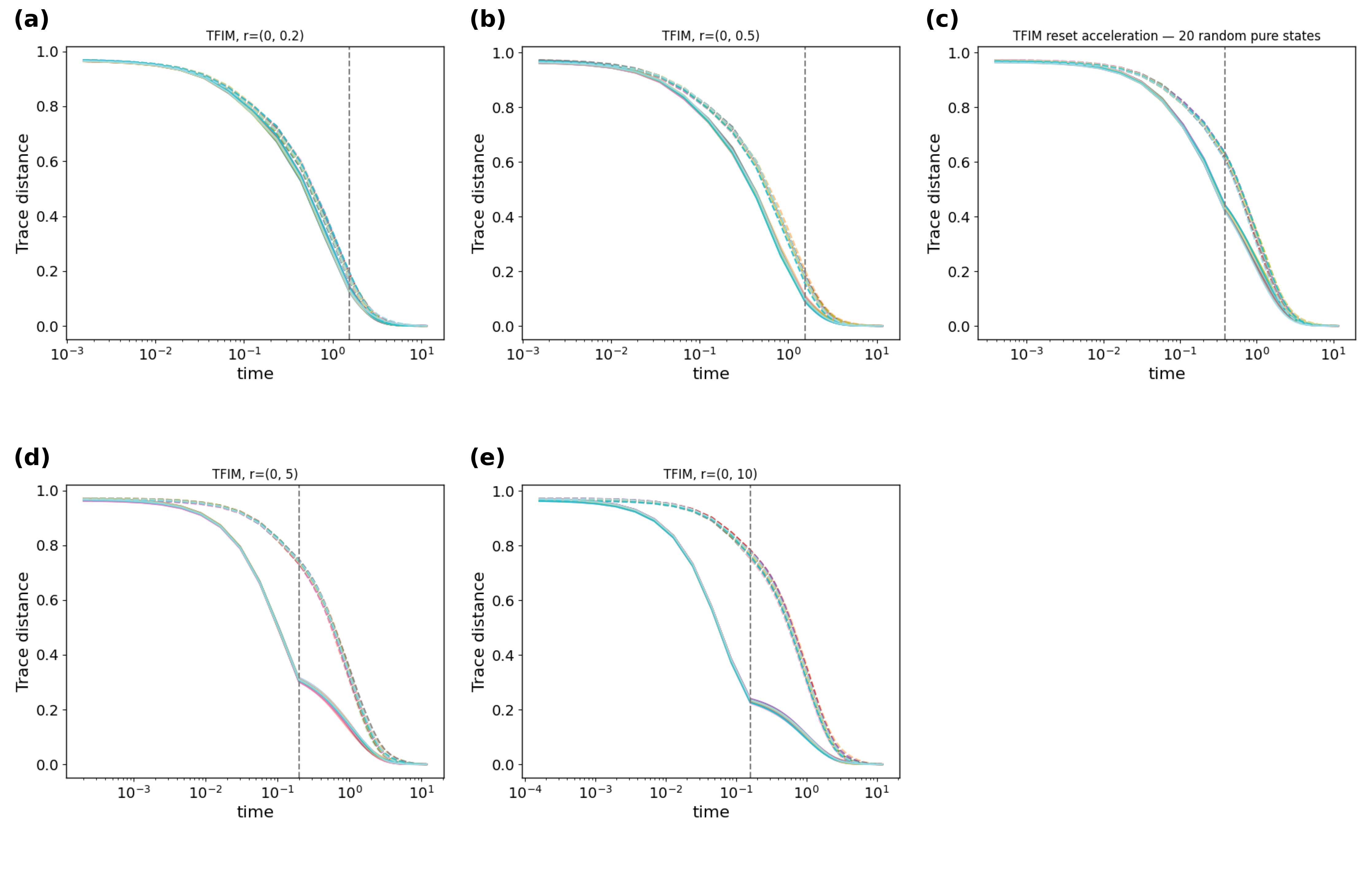}
    \caption{Acceleration of relaxation of the with 20 randomly chosen initial pure states. Dashed lines: without reset; solid: with reset. Parameters: $N=5$, $g=1.2$, $J=\beta=1.0$, $\gamma=0.5$, with different $r$ and $t_s$. (a) $r=0.2$, $t_s=0.8\tau_2$ (b) $r=0.5$, $t_s=0.8\tau_2$ (c) $r=1.0$, $t_s=0.1\tau_2$ (d) $r=5$, $t_s=0.1\tau_2$ (e) $r=10$, $t_s=0.08\tau_2$.}
    \label{S_3}
\end{figure*}

\subsubsection{Robustness of the acceleration under different parameters}
To check the robustness of the maximally mixed state, we numerically examine the fraction of states satisfying the sufficient condition for acceleration presented in the main text, i.e., ${\rm Re}(c_2^*d_2)\leq |c_2|^2$. In Fig. \ref{S_1}, we present the fraction of the initial states among the 2000 randomly picked pure states satisfying the sufficient condition.
\begin{figure}
    \centering
    \includegraphics[width=0.5\linewidth]{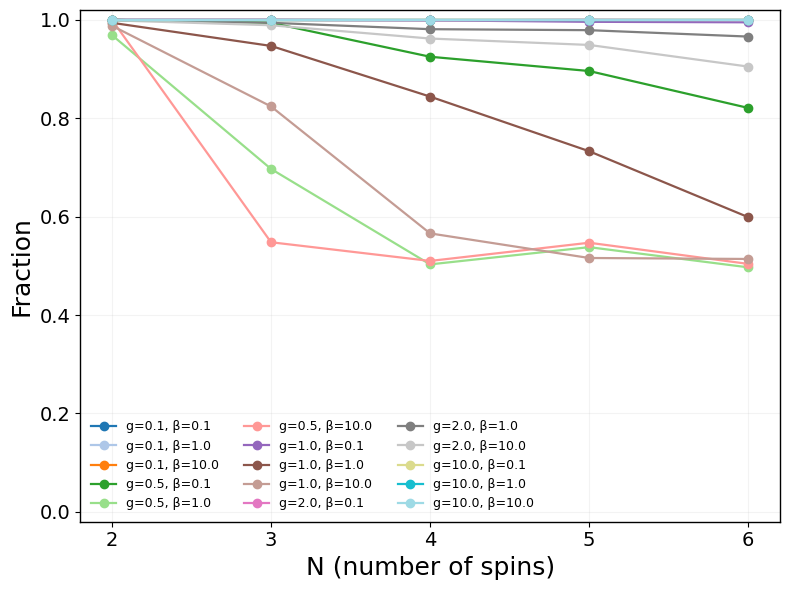}
    \caption{The fraction of states satisfying the acceleration condition, randomly chosen from 2000 pure states. $J=1.0,\ \gamma=0.5$}
    \label{S_1}
\end{figure}

\subsubsection{Simultaneous suppression of multiple relaxation modes with an application to accelerating ``small-gap'' relaxation dynamics}
We calculate the fraction of initial states (among 10000 Haar-random pure states) satisfying multiple sufficient
conditions for suppression of the $k$-th mode, i.e., ${\rm Re}(c_k^*d_k)\leq |c_k|^2$ for different $k$. This is shown in Fig. \ref{S_2}. From this plot, it is clear that multiple such conditions (for different k) can be satisfied simultaneously under various parameters.

\begin{figure}
    \centering
    \includegraphics[width=1.0\linewidth]{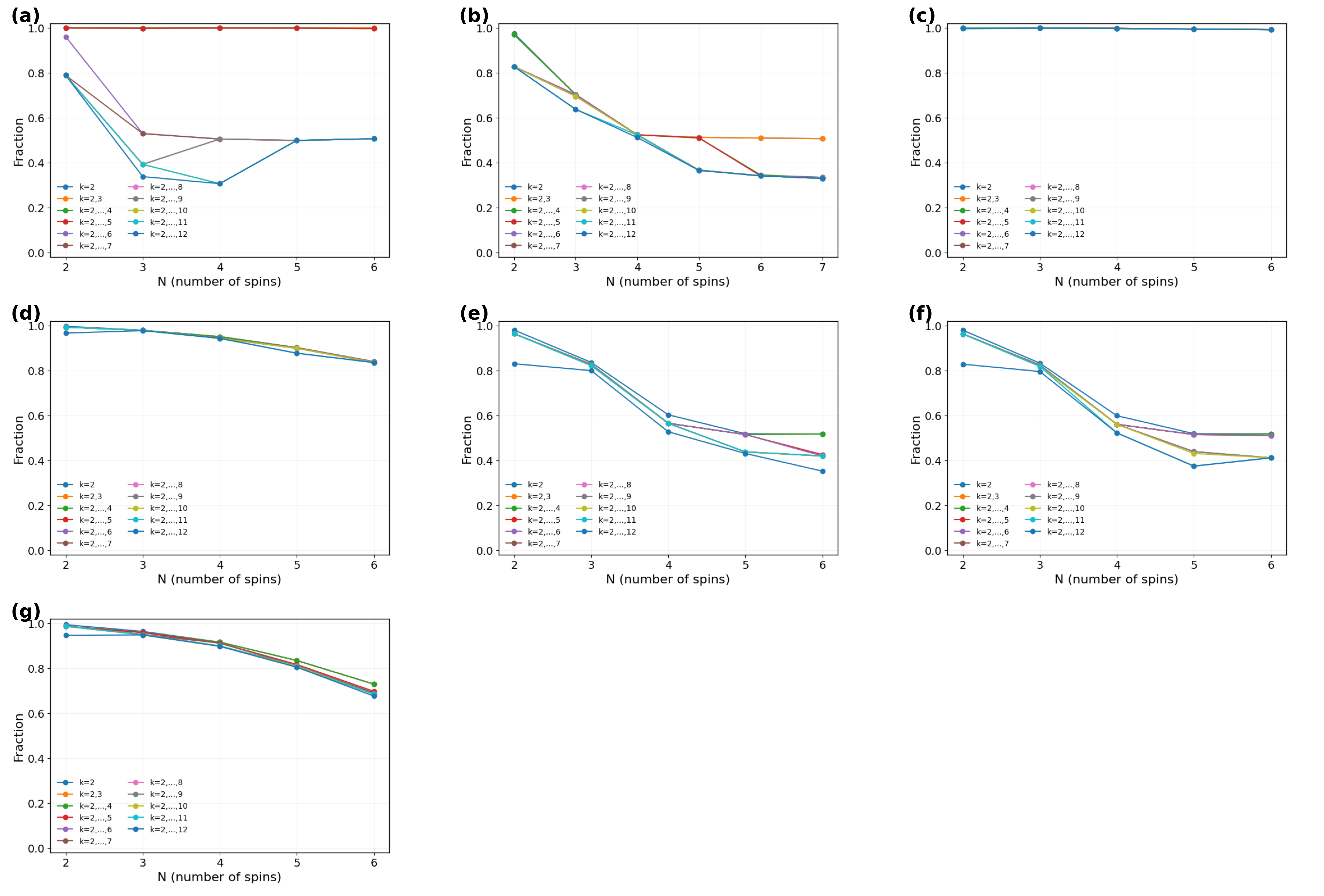}
    \caption{The fraction of states satisfying multiple acceleration condition, randomly chosen from 10000 pure states. $J=1.0, \gamma=0.5$ and different $g$ and $\beta$. For example, the legend ``$k = 2, . . . , 5$'' denotes the simultaneous satisfaction of the conditions ${\rm Re}(c_k^*d_k)\leq |c_k|^2$ for modes $k=2$ through $k=5$. (a) $g=0.1, \ \beta=1.0$ (b) $g=0.5, \ \beta=1.0$ (c) $g=1.0, \ \beta=0.1$ (d) $g=1.0, \ \beta=0.5$  (e) $g=1.0, \ \beta=5.0$ (f) $g=1.0, \ \beta=10.0$  (g) $g=1.2, \ \beta=1.0$.}
    \label{S_2}
\end{figure}

This is particularly useful for cases where the first few non-zero eigenvalues are very close, making the suppression of only the slowest mode insufficient for acceleration. For illustrative purposes, we consider a specific case with $g=0.1$, $\beta=1.0=J$, $\gamma=0.5$. In this example, the real parts of the first four nonzero eigenvalues are very close: $\lambda_2\approx -0.5534+1.987i$, $\lambda_3^{*}\approx-0.5534-1.987i$, $\lambda_4\approx-0.5545+1.987i$, and $\lambda_5^{*}\approx-0.5545-1.987i$, with a clearer gap at $|{\rm Re}\,\lambda_6|=0.685$. As Fig. \ref{S_4} shows, acceleration persists despite the small gap, consistent with the large fraction of random states satisfying the multiple suppression conditions (for $k=2$-$5$) simultaneously. 

\begin{figure}
    \centering
    \includegraphics[width=0.8\linewidth]{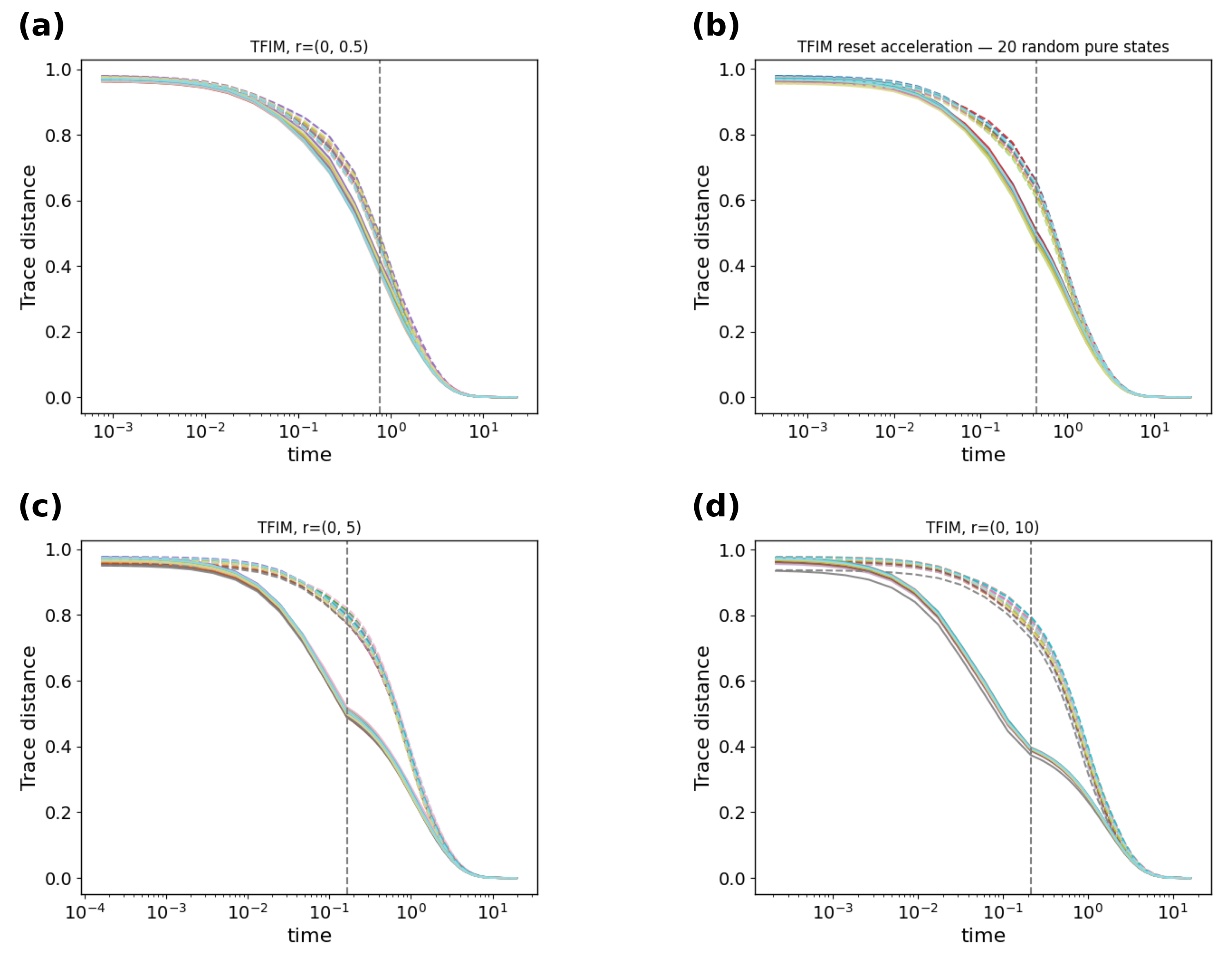}
    \caption{Acceleration of relaxation in the TFIM for the small-gap case, using 20 randomly chosen initial pure states. Dashed lines: evolution without reset; solid lines: with reset. Parameters: $N=5$, $g=0.1$ (small spectral gap), $J=\beta=1.0$, $\gamma=0.5$, with varying reset rate $r$ and reset interval $t_s$.
    (a) $r=0.5$, $t_s=0.2\tau_2$ (b) $r=1.0$, $t_s=0.1\tau_2$ (c) $r=5.0$, $t_s=0.1\tau_2$ (d) $r=10.0$, $t_s=0.08\tau_2$}
    \label{S_4}
\end{figure}
}

\clearpage
\bibliography{bibfile}